\documentclass[%
 reprint,
 amsmath,amssymb,
 aip, 
]{revtex4-2}

\usepackage{placeins}
\usepackage{afterpage}

\usepackage{graphicx}% Include figure files
\usepackage{dcolumn}% Align table columns on decimal point
\usepackage{bm}% bold math
\usepackage{hyperref}% add hypertext capabilities

\usepackage{libertinus}

\begin{document}

%\preprint{APS/123-QED}

\title{Finite-temperature micromagnetic model bridging atomic- and macro-scale magnetism}% Force line breaks with \\

%\thanks{A footnote to the article title}%

\author{R. Kiefe}
 \email{rodrigokiefe@ua.pt}
 \affiliation{Departamento de Física and CICECO, Universidade de Aveiro, 3810-193 Aveiro, Portugal}

\author{J. S. Amaral}
\affiliation{Departamento de Física and CICECO, Universidade de Aveiro, 3810-193 Aveiro, Portugal}

%\collaboration{MUSO Collaboration}%\noaffiliation

\date{\today}

\begin{abstract}

A multi-scale finite-temperature micromagnetic model is presented, based on the Landau-Lifshitz equation and the Bernoulli differential equation. This model accurately reproduces classic Maxwell magnetostatics of paramagnets for high temperatures and accurately reproduces standard micromagnetics described by the conventional Landau-Lifshitz model in ferromagnets. The Landau-Lifshitz-Bernoulli (LLBe) model can, by design, directly couple atomic-scale simulations with micromagnetics and output consistent predictions of bulk magnetic properties at finite temperatures, from below to above the material's Curie temperature. The LLBe model is validated against established solvers: MUMAX3 for zero-temperature micromagnetics, and FEMCE for high-temperature classic magnetostatics. We present two examples of applications of the LLBe model: Heat-Assisted magnetic recording on a thin magnetic track with local heating; and temperature-dependent coercivity in Nd\textsubscript{2}Fe\textsubscript{14}B magnets, demonstrating the multi-scale finite-temperature simulation capabilities of the LLBe.

\end{abstract}


%Use showkeys class option if keyword
%display desired

\maketitle

\section{Introduction} \label{sec:intro}

	% Add some of these references:
		% https://www.nature.com/articles/s44306-026-00139-x
		% https://www.nature.com/articles/s41524-025-01893-y
		% https://www.nature.com/articles/s42256-024-00914-7
		% https://www.nature.com/articles/s41524-026-02174-y
		% https://www.nature.com/articles/s44306-026-00134-2
		% https://www.nature.com/articles/s41524-025-01926-6
		% https://www.nature.com/articles/s41524-025-01688-1
		% https://www.nature.com/articles/s41524-022-00719-5
		
	 % === High temperature Neel skyrmions ===
	 % They use mumax3, scale Ms(T) = Ms(0) (1-(T/Tc)^a)
	 % with Ms(0) = 598 kA/m, a = 1.92, Tc = 1216 K
		% https://www.nature.com/articles/s41467-026-73775-w

	 % Laser induced nucleation
		% https://www.nature.com/articles/s41567-026-03236-0

	The performance of magnetism-based technologies, from permanent magnets in wind turbine generators and traction motors \cite{OG_2011} to soft ferromagnets in transformer cores \cite{OG_2011} and ferromagnetic thin films in hard disk drives, is fundamentally limited by the properties of the employed magnetic materials. These applications demand distinct characteristics: high coercivity for permanent magnets, high permeability and low heat loss for soft magnetic components, and high read/write speeds and thermal stability for data storage media \cite{Challener2009}. Computational modeling has therefore become essential to accelerate materials development. However, the fidelity of such models depends critically on modeling the relevant physical behavior, from atomic-scale exchange interactions to micro-scale domain wall dynamics and macro-scale shape effects, at the application-relevant length scale. 

	Micromagnetics describes magnetism for nano- and micro-scale ferromagnetic materials, based on the phenomenological Landau-Lifshitz equation \cite{Leliaert_2019} (LL), here written as follows \cite{Oriano_2008},

% Landau-Lifshitz
\begin{equation}
\begin{split} \label{eq:LL}
	\frac{1}{\gamma} \frac{\partial \vec{M}}{\partial t} = &- \vec{M} \times \vec{H} \\ 
		&- \frac{\alpha}{M_s} \vec{M} \times \left( \vec{M} \times \vec{H} \right) \\
		\vec{H} = & \vec{H}_a + \vec{H}_d + \vec{H}_{an} + \vec{H}_{ex} + ...,
\end{split}
\end{equation}

	\noindent where $\gamma$ is the gyromagnetic ratio, $\alpha$ is a damping constant, $M_s$ is the saturation magnetization and $\vec{H}$ is the effective field - a sum of each field contribution, usually defined by an applied field $H_a$, the demagnetizing/magnetostatic field $H_d$, the exchange field $H_{ex}$ and an anisotropy field $H_{an}$ \cite{Oriano_2008,Mumax3_2018}. Additional terms can be considered, such as the Dzyaloshinskii–Moriya interaction \cite{Moriya_1960} and thermal noise (Langevin field) \cite{Garanin_1997}.

	The LL equation considers a continuous magnetization field $\vec{M}$, with fixed magnitude $M_s$, which is equivalent to stating that the temperature of the magnetic volume is much lower than its critical temperature and is fully saturated \cite{Leliaert_2019}. In Eq. \ref{eq:LL}, the first cross product $\vec{M}\times\vec{H}$ rotates the magnetization around the effective field $H$, and the second cross product $\vec{M} \times \left(\vec{M} \times \vec{H}\right)$ relaxes the magnetization towards the effective field. This dynamic equation minimizes the Gibbs free energy of the system until $\vec{M} \times \vec{H} = 0$ \cite{Scholz_2003}.

	The fixed norm, zero-temperature assumption of the LL equation limits its scope and applications in modelling magnetic materials. Finite-temperature micromagnetic models are essential when studying dynamic, high temperature systems such as Heat-Assisted Magnetic Recording (HAMR) \cite{Leliaert_2019}, the calculation of the mobility of domain walls in rare-earth ferrite garnets \cite{Garanin_1997} and the non-linear thermal response of magnetocaloric materials under applied magnetic field changes \cite{Basso_2024,Ohmer_2020}. 

	Several methods have integrated temperature to micromagnetics, such as by introducing a stochastic thermal field to the LL equation (sLL) \cite{Skomski_2013,Wei_2016,Atxitia_2017,Oezelt2022}. However, the sLL is inaccurate at temperatures close-to or above the Curie temperature ($T_c$) of the material \cite{Wei_2016,Atxitia_2017,Leliaert_2019,Lepadatu_2021}. Another strategy is to re-normalize the input parameters of the LL equation for the target temperature \cite{Skomski_2013,Wei_2016,Gija_2025}, but this approach still locks the magnetization magnitude to a fixed value and assumes a spontaneous magnetization above $T_c$ which is not physical \cite{Atxitia_2017,Gija_2025}. The sLL, in combination with parameter re-normalization, was recently successfully employed to model Néel skyrmions in Co-Al and Co-Ni-Al \cite{Wang2026} at elevated temperatures up to 773 K, approximately 440 K below $T_c$ - within the temperature limit of this approach.

	The magnetization norm, as described by the LL equation, is always preserved in principle, by the equality $\partial |\vec{M}|^2 / \partial t = 2 \vec{M} \cdot \partial \vec{M} / \partial t = 0$ \cite{Visintin1985}, which is a limiting description at the micro-scale, as discussed above. The Landau-Lifshitz-Bloch (LLB) equation approaches this limitation by allowing for the relaxation of the magnetization length \cite{Atxitia_2017}. This model includes the nonlinear response of the local magnetization to the local effective field $H$ \cite{Garanin_1997,Atxitia_2017}, written as

% Landau-Lifshitz-Bloch
\begin{equation}
\begin{split} \label{eq:LLB}
		\frac{1}{\gamma} \frac{\partial \vec{M}}{\partial t} = &- \vec{M} \times \vec{H} +  \frac{\alpha_\parallel}{|\vec{M}|^2} \left( \vec{M} \cdot \vec{H}  \right) \vec{M} \\ 
		&- \frac{\alpha_\perp}{|\vec{M}|^2} \vec{M} \times \left( \vec{M} \times \vec{H} \right),
\end{split}
\end{equation}

	\noindent where $\alpha_\parallel$ and $\alpha_\perp$ are the longitudinal and transverse damping coefficients and the effective field is similar to the conventional effective field in Eq. \ref{eq:LL}, but with an additional term that differs on whether the material's temperature is above or below $T_c$ \cite{Garanin_1997,Atxitia_2017}, proportional to the longitudinal and transversal magnetic susceptibilities $\chi_\parallel$ and $\chi_\perp$. An immediate limitation of the LLB is that it requires information on both the transverse and longitudinal magnetic response of the material both in the form of damping coefficients and individual susceptibilities. Also, it is not straight-forward to include a predetermined temperature and magnetic field dependence of the magnetization - $M(H,T)$ - in the LLB, since it is often parameterized by zero-field properties of the material \cite{Vogler2014,Moretti2017}.

\section{A new micromagnetic model}

	To overcome the limitations of conventional micromagnetics described by the LL and LLB models, we present a new micromagnetic model, based on the Landau-Lifshitz Eq. \ref{eq:LL} and the Bernoulli type of differential equations, as follows:

% LLBe
\begin{equation}
\begin{split} 	\label{eq:LLBe}
	\frac{1}{\gamma} \frac{\partial \vec{M}}{\partial t} = &- \vec{M} \times \vec{H} \\ 			&- \alpha \, \vec{M} \times \left( \vec{M} \times \vec{H} \right) \\
				 &- \beta\left(|\vec{M}| - \mathcal{M}(H, T) \right) \vec{M}
\end{split}
\end{equation}

	This equation is inspired by the Callen equation \cite{Callen1958,Berger_2000}, where the longitudinal diffusion in Eq. \ref{eq:LLBe} is redefined as a function of the equilibrium magnetization state $\mathcal{M}$, which depends on the local effective field $H$, and its temperature $T$, as opposed to being a constant.

	Just like with the LL equation, this Landau-Lifshitz-Bernoulli (LLBe) equation minimizes the energy towards an equilibrium state $\vec{M} \times \vec{H} = 0$, while the magnetization norm is no longer fixed. When this equilibrium condition is reached, Eq. \ref{eq:LLBe} is reduced to a Bernoulli differential equation of degree 2, which represents a logistic growth - the magnetization norm grows/shrinks towards the equilibrium value $\mathcal{M}(H,T)$, which itself depends on the local effective field. This equilibrium data can come from experimental measurements, mean-field theory (MFT) or atomic-scale models such as the Heisenberg or Ising models. % With this approach, the LLBe has direct information on the material's intrinsic magnetization and a straight-forward computational implementation structure in mind.

	Unlike the Landau-Lifshitz-Bloch equation, the proposed LLBe model does not add additional field contributions to the effective field. The LLBe keeps the standard expressions while replacing $M_s$ with the local magnetization norm $|\vec{M}|$,

% Magnetic field contributions (FEM)
\begin{equation} \label{eq:Hcontributions}
	\vec{H} = \vec{H}_a + \vec{H}_d + \frac{2 k_{an}}{\mu_0 |\vec{M}|^2} \left(\vec{M} \cdot \vec{u}  \right) \vec{u} + \frac{2 k_{ex}}{\mu_0 |\vec{M}|^2} \nabla^2 \vec{M} + ...
\end{equation}

	\noindent where $k_{an}$ and $k_{ex}$ are the anisotropy and exchange energy terms - a unique expression whether below or above $T_c$. Special care must be taken to make sure that the underlying energy functional of each magnetic field contribution is consistent with this new description. We verify the validity of Eq. \ref{eq:Hcontributions} using the variational calculus approach described in [\onlinecite{Scholz_2003}], ensuring that the proposed magnetic field expression remains consistent to the respective energy functional. % A brief overview of this validation step is presented for the exchange field in the next section.

	% Particularly, the original anisotropy energy $E_{an} = \int_\Omega k_{an} \left( 1-(\vec{u}_{an} \cdot \vec{m}) \right)^2 \, dV$ \cite{Scholz_2003} is no longer accurate. Instead, the system's uniaxial crystal anisotropy energy can be described by \cite{Visintin1985} 

% \begin{equation*}
% 	E_{an} = \int_\Omega \frac{k_{an}}{M_s^2} \left( M^2 - \left( \vec{u}_{an} \cdot \vec{M} \right)^2  \right) \, dV .
% \end{equation*}

% 	\noindent The anisotropy field corresponding to this energy functional, 

% \begin{equation*}
% 	\vec{H}_{an} = -\frac{2k_{an}}{M_s^2} \left( \vec{M} - \left( \vec{M} \cdot \vec{u}_{an}  \right)\vec{u}_{an} \right),
% \end{equation*}

% 	\noindent can be simplified to the more familiar anisotropy field expression in (\ref{eq:Hcontributions}), since $\vec{H}$ always appears as a cross product with $\vec{M}$.

\subsection{Numerical implementation}

	In this work, the time derivative of the LLBe equation was discretized by the Finite Difference Method, while the magnetic field contributions and the magnetization field were discretized following the Finite Element Method. In Eq. \ref{eq:Hcontributions}, only the demagnetizing field and the exchange field need to be estimated by a numerical method, as the other field contributions can be computed directly. The demagnetizing field is obtained following a hybrid FEM-BEM approach\cite{Oriano_2008} and the exchange field is obtained by the box method \cite{Scholz_2003}. The time step procedure is inspired by Bottauscio et al. \cite{Oriano_2008}, adapted considering the LLBe's extra term.

\section{Low temperature validation of the LLBe}

	% A LLBe consegue simular micromagnetismo convencional, descrito pela LL usual
	% E consegue simular magnetoestatica classica de Maxwell

	When a ferromagnetic material is at a temperature far below $T_c$, the equilibrium magnetization $\mathcal{M}$ is the saturation magnetization $M_s$ and the LLBe naturally recovers the conventional Landau-Lifshitz equation. In this low temperature setting, the LLBe should replicate existing results from standard micromagnetics software such as OOMMF \cite{OOMMF} and MUMAX3 \cite{Mumax3_2014}.

\begin{figure}[htp]
	\centering
	\includegraphics[width=0.9\linewidth]{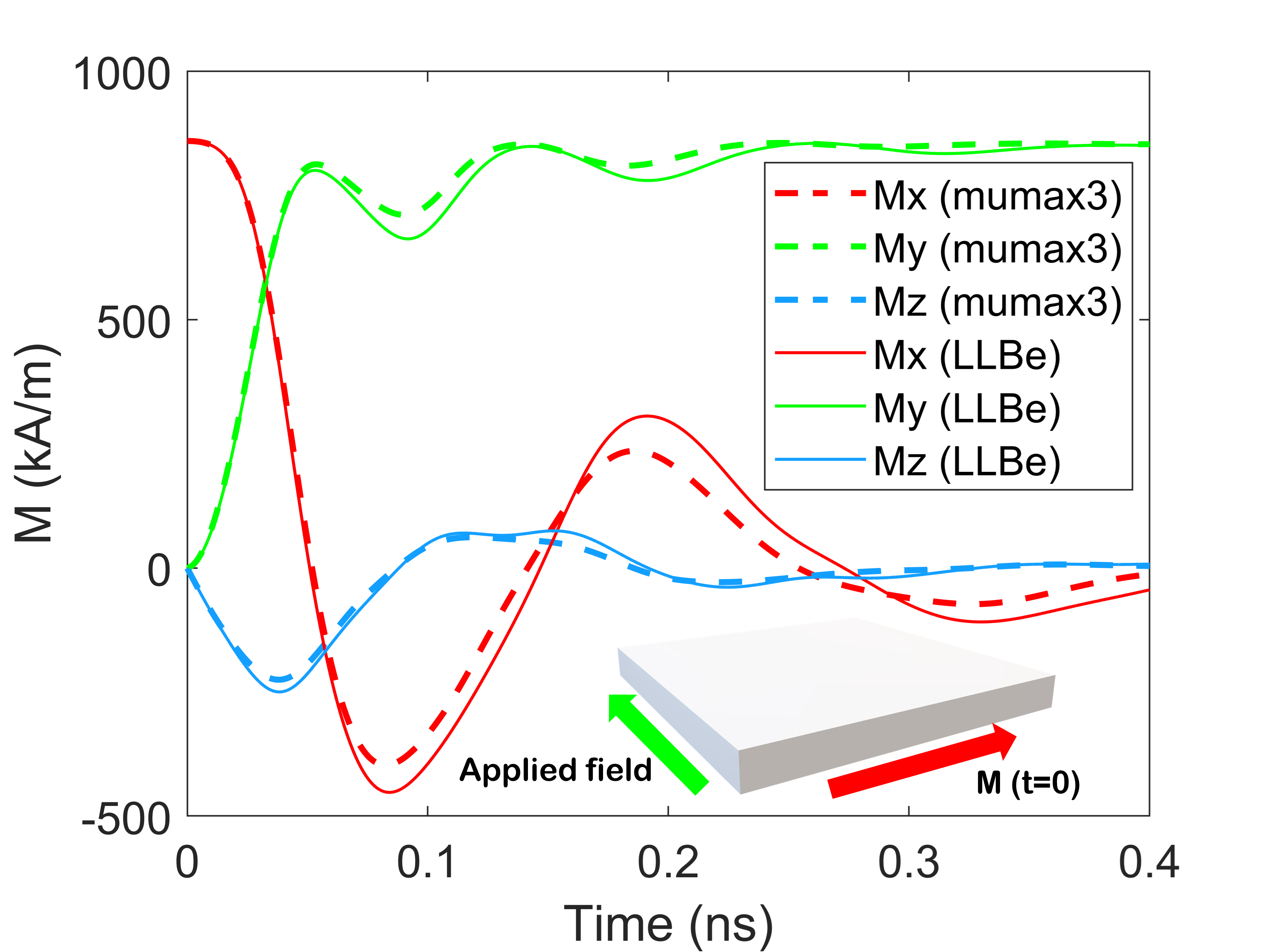}
	\caption{LLBe and MUMAX3 simulation of a permalloy thin-film's average magnetization over time.}
	\label{fig:permalloy}
\end{figure}

	We use MUMAX3 to replicate the micromagnetic simulation of a permalloy thin film by Bottauscio et al. \cite{Oriano_2008} and then compare the results to those of the LLBe. The thin film has the dimensions of 100 x 100 x 5 nm, with model parameters $M_s = $ 860 kA/m, $k_{ex} = $ 13 pJ/m, modeled with $\alpha = $ 0.1 and a fixed time-step $\Delta t = $ 67 fs. The initial magnetization is uniform and along the $x$-axis, with an applied field in the $y$-axis of 50 kA/m. As shown in Figure \ref{fig:permalloy}, the dynamic behavior of the x,y,z components of the average magnetization of the permalloy as predicted by MUMAX3 is quantitatively well reproduced by the LLBe. 

	% The C++ numerical implementation for this simulation is available at \url{codeberg.org/rkiefe/Femeko.jl/src/branch/main/Femeko.cpp/Micromagnetics}, and took 21.28 seconds to compute, running on an Intel i5-14400. The computational performance of the numerical method is expected to be drastically improved with more sophisticated methods, but this is outside the scope of this work.

	% !!! NOTE: the permalloy simulation in C++ toke 21.281 seconds for a mesh with 8137 elements in total and 865 elements in the magnetic region running on an Intel i5-14400.

% LLBe vs Oriano 2008 (OOMMF) | Permalloy

\section{Validation of the LLBe for T near T\textsubscript{C}}

	A relevant feature of the LLBe is that when considering only the demagnetizing and applied field contributions to the effective field, the LLBe simulation of the minimum energy state is equivalent to one obtained by classic Maxwell magnetostatic solvers. To illustrate this, we consider a cube of metallic Gd (a ferromagnet with T\textsubscript{C} $\approx$ 293 K), whose $\mathcal{M}(H,T)$ input data was obtained from Weiss molecular mean-Field theory. The spatially-resolved magnetization was simulated with both FEMCE \cite{Kiefe_2025} (a 3D finite element simulation software for magnetocalorics) and the LLBe, for temperatures slightly below and above T\textsubscript{C}.

\begin{figure}[htp]
	\centering
	\includegraphics[width=0.39\linewidth]{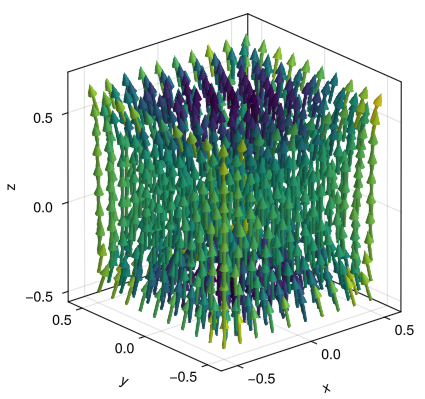}
	\includegraphics[width=0.51\linewidth]{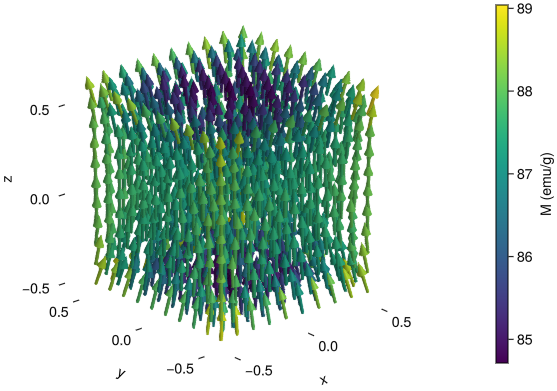}
	\caption{Minimum energy magnetostatic (left) and LLBe (right) simulation of a Gd cube magnetization field $\vec{M}$ at 280 K, under an applied field of 1.0 T along the $z$-axis.}
	\label{fig:LLBe_vs_FEM_280K}
\end{figure}

\begin{figure}[htp]
	\centering
	\includegraphics[width=0.38\linewidth]{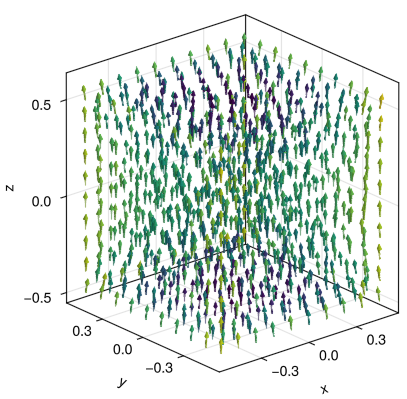}
	\includegraphics[width=0.52\linewidth]{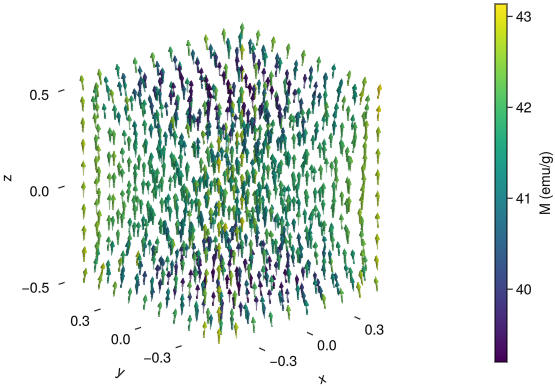}
	\caption{Minimum energy magnetostatic (left) and LLBe (right) simulation of a Gd cube magnetization field $\vec{M}$ at 300 K, under an applied field of 1.0 T along the $z$-axis.}
	\label{fig:LLBe_vs_FEM_300K}
\end{figure}

	Both FEMCE and the LLBe predict a magnetization field of a 1 x 1 x 1 cm cube at 280 K and at 300 K ($\sim$ 10 K below and above Tc) under an applied field of 1.0 Tesla along the $z$-axis (see Figures \ref{fig:LLBe_vs_FEM_280K} and \ref{fig:LLBe_vs_FEM_300K}) with good quantitative agreement, achieving a maximum relative difference of 0.06\% between the two models. For this simulation, the LLBe coefficients were set as $\beta = 2.0$ and $\alpha = 1.0$ to speed up the convergence rate towards equilibrium, and every remaining energy term was set to 0. The LLBe solver is stopped when $|\partial \vec{M}/ \gamma \partial t| < 10^{-6}$. 

% FEM vs LLBe simulation of Gd cube at 280 K

% FEM vs LLBe simulation of Gd cube at 300 K

	This LLBe model demonstrates an accurate prediction of both classical macro-scale magnetism and conventional micromagnetics, with a single equation for both above and below $T_c$ - a suitable model for dynamic multi-scale high-temperature magnetic systems such as in heat-assisted magnetic recording as exemplified in the next section. 
	
% \FloatBarrier
\section{Application: Heat Assisted Magnetic Recording} \label{sec:HAMR}

	In the scope of computer data storage, Heat-Assisted Magnetic Recording (HAMR) is a promising technology in increasing storage density \cite{Leliaert_2019, Ababei_2019, Meo_2020, Gija_2025}. The underlying magnetic recording strategy relies on materials with grains of high crystalline anisotropy and a heating element (a laser) to locally heat the material close to $T_c$ to reduce its coercivity - decreasing the write field strength required \cite{Meo_2020}. Conventional micromagnetic models struggle to model this recording process, due to Finite-Temperature micromagnetics being a multi-scale temperature dependent process \cite{Meo_2020,Gija_2025}. The LLBe brings a direct bridge between thermomagnetic properties and nano-scale simulations, making it a suitable approach for modelling HAMR. % powerful/suitable

	To show this, we consider a single-layer magnetic track similar to the study reported by Gija et al. \cite{Gija_2025}: 540 x 60 x 16 nm, a fixed value of the exchange parameter $k_{ex} = $ 6.8e-12 J/m and a zero-temperature $k_{an, \, 0} = $ 2.2e6 J/m$^3$ which is then scaled by $k_{an} (T) = k_{an, \, 0} \left[ \frac{\mathcal{M}(H=0,T)}{M_s} \right] ^\eta$, with $\eta = 3$, a reasonable value for HAMR media \cite{Wei_2016}. The easy-axis magnetization is in the $z$-axis, perpendicular to the strip. The equilibrium temperature and magnetic field dependent magnetization data $\mathcal{M}$ was obtained via a recent Mean-Field model with crystalline anisotropy \cite{Klunnikova2025}, considering a T$_C = $ 710 K.

% Figure of M data of the track
% \begin{figure}[htp]
% 	\centering
% 	\includegraphics[width=0.9\linewidth]{M_HAMR_data.png}
% 	\caption{Equilibrium input magnetization data $\mathcal{M}(H,T)$ for the HAMR simulation with the LLBe model - eq. (\ref{eq:LLBe}) - obtained via a recent Mean-Field model with uniaxial anisotropy\cite{Klunnikova2025}.}
% 	\label{fig:hamr_data}
% \end{figure}

\begin{figure}[htp]
		\centering
		\includegraphics[width=.9\linewidth]{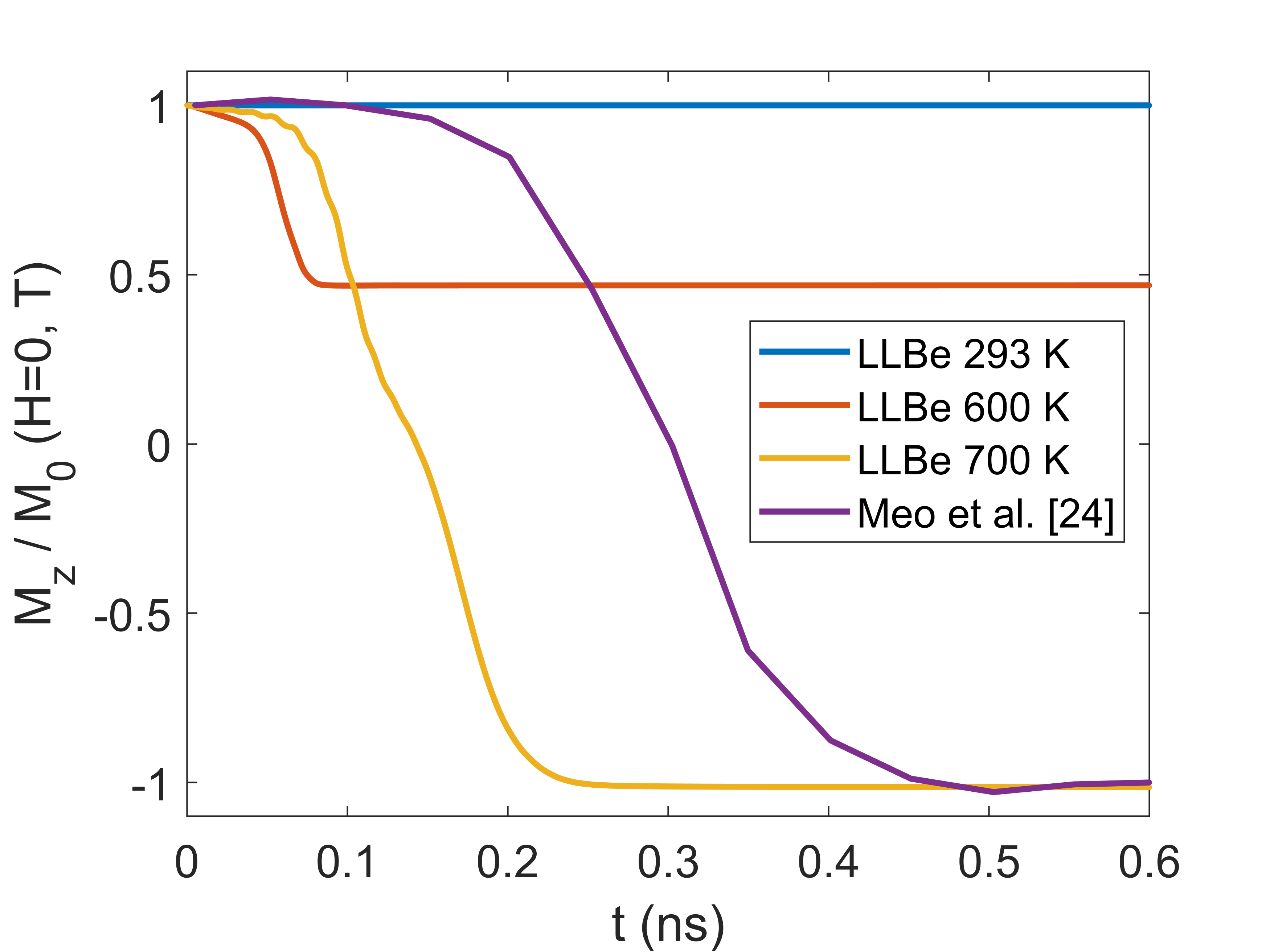}
		\caption{LLBe simulation of the heat-assisted recording of a bit on a magnetic strip (540x60x16 nm), showing a dynamic switching behavior consistent with other existing material simulations \cite{Meo_2020}.}
		\label{fig:HAMR_LLBe}
	\end{figure}

	Each bit (25 nm wide) of the magnetic track is at room temperature (293 K), with its magnetization pointing downwards in the $z$-axis, except for the bit at the center of the track which points up. A uniform, local external field of 3 T is applied on that bit, opposing its magnetization. The dynamic flip of the bit's magnetization is modeled by the LLBe equation, for 3 bit temperatures: 293 K, 600 K, and 700 K (10 K below $T_c$). For this applied field strength, the bit's average $M_z$ component is flipped successfully only when assisted by a local temperature increase to 700 K (see Figure \ref{fig:HAMR_LLBe}). Without this temperature increase, the bit's magnetization is not inverted, as expected. This phenomenon was verified for different materials and magnetic track configurations \cite{Meo_2020,Gija_2025,Ababei_2019}, with bit flip rates qualitatively consistent with LLBe simulations. The LLBe does not reproduce the exact bit flip rates declared in \cite{Meo_2020}, but this is to be expected as they consider a different material (FePt) and simulation approach (described as a granular model based on the stochastic LLB equation).

% Figure of Mz(t) simulated with LLBe

\section{Application: temperature-dependent coercivity in Nd-Fe-B magnets}

Nd-Fe-B magnets are one of the strongest and most energy-dense magnets available. In some key applications, such as in the automotive industry, they must operate at elevated temperatures approaching T$_C$ ($\sim$ 600 K). At this temperatures the coercive field H$_C$ decreases sharply and the magnet eventually self-demagnetizes - a long standing material challenge known as ``\textit{the coercivity problem} of Nd-Fe-B'' \cite{Okamoto2015}. Understanding and predicting the temperature dependence of H$_C$ is therefore of direct technological relevance.

Conventional micromagnetics are not suitable to study the temperature dependence of H$_C$ in these materials for the reasons discussed in Sections \ref{sec:intro} and \ref{sec:HAMR}: the fixed-norm assumption of the LL equation breaks down close to T$_C$ and parameter renormalization does not physically describe the system. The LLBe model incorporates the equilibrium magnetization $\mathcal{M} (H, T)$ directly into the governing equation, making it a natural framework for simulating hysteresis loops across a wide temperature range with a single set of standard-literature material parameters.

\begin{figure}[ht]
	\centering
	\includegraphics[width=\linewidth]{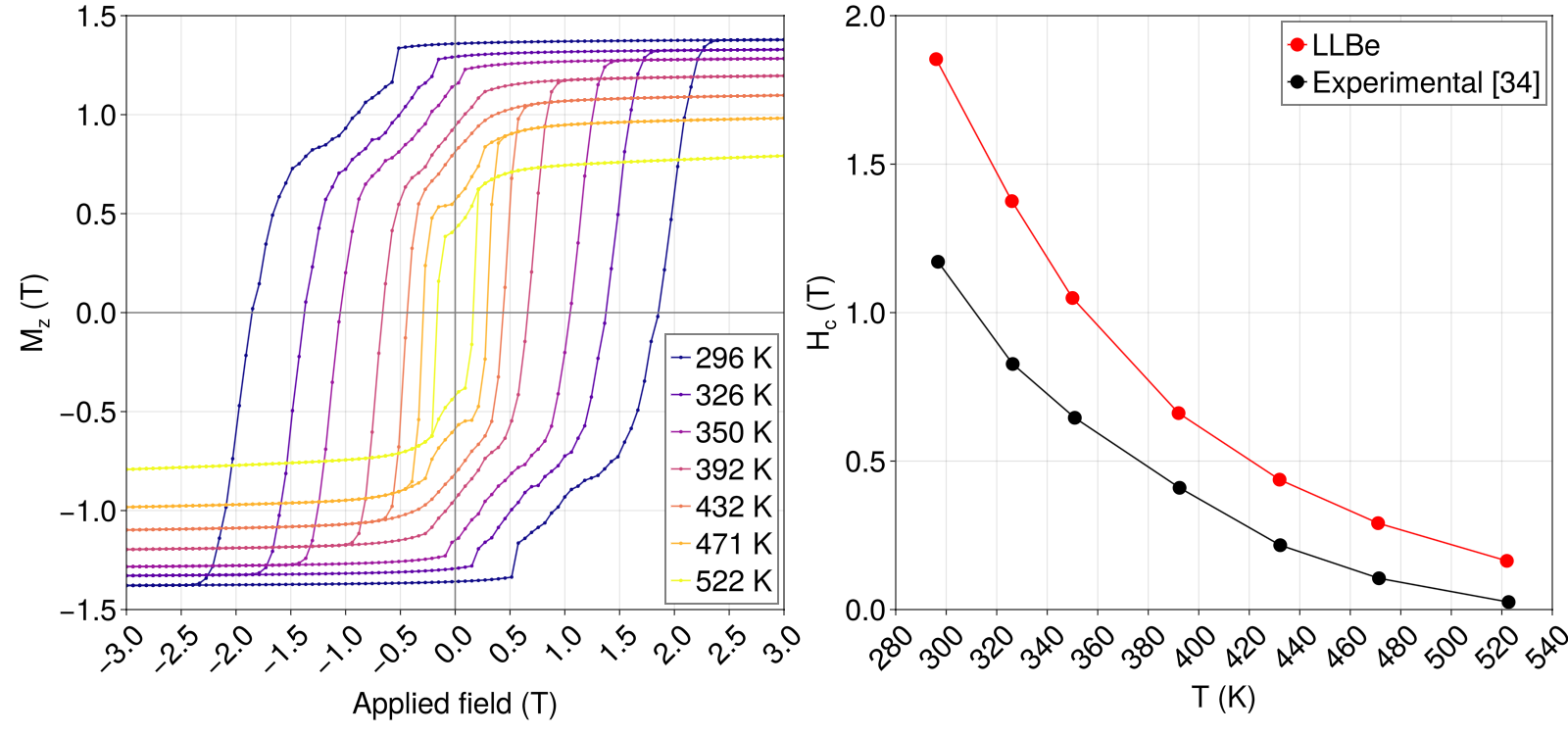}
	\caption{LLBe simulation of temperature-dependent hysteresis loops (left) and the temperature dependence of the coercive field (right) of a single-crystalline cube of Nd\textsubscript{2}Fe\textsubscript{14}B.}
	\label{fig:NdFeB}
\end{figure}

To illustrate this capability, we simulate the hysteresis loop of a 1 x 1 x 1 cm single-crystalline cube of Nd\textsubscript{2}Fe\textsubscript{14}B at temperatures between 300 K and 523 K, and compare directly with experimental measurements. The model parameters are taken from standard literature values\cite{coey_book,Fischbacher2018,Lars1991}: a zero-temperature saturation magnetization $M_s$ of 1.61 T, a zero-temperature anisotropy constant $K_{an}$ of 4.9 MJ/m\textsuperscript{3}, T$_C$ at 588 K and angular momentum J = 9/2. The anisotropy constant is scaled for different temperature by the same expression as in Section \ref{sec:HAMR}, with $\eta = 10$ a conventional coefficient for cubic crystals \cite{Akulov_1936}. The equilibrium magnetization $\mathcal{M}(H, T)$ is estimated via the Weiss molecular mean-field theory. The LLBe estimate of $H_C (T)$ has surprisingly good agreement with experimental measurements\cite{Sagawa1987} (see Figure \ref{fig:NdFeB}) given the approximations considered.

This result demonstrates that the LLBe model is able to capture the temperature dependence of the coercive field in a technologically relevant permanent magnet, using a single governing equation and standard input data from the literature. This is a capability that conventional micromagnetic solvers do not offer, and positions the LLBe as a practical tool for the multi-scale modelling of finite-temperature magnetic materials.

\FloatBarrier
\section{Conclusion}

A new finite-temperature micromagnetic model is presented, capable of simulating magnetic materials from below to above the material's Curie temperature, with a single governing expression - the Landau-Lifshitz-Bernoulli (LLBe) equation - inspired by the Callen equation \cite{Callen1958}. This LLBe model couples equilibrium magnetization data $M(H,T)$ (obtained from either atomistic models or experimental measurements) to micromagnetic simulations, and smoothly recovers conventional zero-temperature micromagnetic results and classic Maxwell magnetostatics.

With the proposed LLBe model we have estimated the temperature dependency on the coercive field in a Nd\textsubscript{2}Fe\textsubscript{14}B single-crystal, with good agreement with experimental measurements reported in the literature. The LLBe model was also capable of predicting memory write speeds and their dynamic process in heat-assisted magnetic recording media, replicating the behavior reported in a previous study\cite{Meo_2020} on a different material (FePt) using a multiscale approach based on the Landau-Lifshitz-Bloch equation.

In summary we have demonstrated that the proposed LLBe model can replicate relevant finite-temperature magnetic properties (spin-flip rates and the coercive field dependence on temperature) in diverse applications, with a versatile, deterministic approach, which significantly broadens the range of potential micromagnetics-based research.

% This novel finite-temperature micromagnetic model allows the multi-scale computational modeling of magnetocalorics and other temperature-critical magnetic materials, extending previous micromagnetic studies in La-Fe-Si \cite{Lai2021} and Co$_2$B \cite{Ohmer_2020} materials. Future work should apply the LLBe model to these and related systems to establish guidelines on how to improve their performance.

% With this novel finite-temperature micromagnetic model, a new pathway for computational modeling of the role of the microstructure in magnetocaloric materials is presented, boosting the search and eventual discovery of key performance gains for magnetic cooling.

\begin{acknowledgments}

This work was developed within the scope of the project CICECO Aveiro Institute of Materials, UID/50011/2025 (DOI 10.54499/UID/50011/2025) \& LA/P/0006/2020 (DOI 10.54499/LA/P/0006/2020), financed by national funds through the FCT/MCTES (PIDDAC). This work has received funding from the European Union’s Horizon Europe research and innovation program through the European Innovation Council (Grant Agreement No. 101161135—MAGCCINE). R. Kiefe acknowledges FCT for PhD grant reference 2025.02860.BD.

\end{acknowledgments}

%\appendix
%\section{Appendixes}

\section*{References}
\bibliography{biblio.bib}

\end{document}